# Single-Hemisphere Photoelectron Momentum Microscope with Time-of-Flight Recording

G. Schönhense*, S. Babenkov, D. Vasilyev, H.-J. Elmers and K. Medjanik

*Johannes Gutenberg-Universität, Institut für Physik, 55128 Mainz, Germany*
* corresponding author: schoenhe@uni-mainz.de

## Abstract

Photoelectron momentum microscopy is an emerging powerful method for angle-resolved photoelectron spectroscopy (ARPES), especially in combination with imaging spin filters. These instruments record $k_x$-$k_y$ images, typically exceeding a full Brillouin zone. As energy filters double-hemispherical or time-of-flight (ToF) devices are in use. Here we present a new approach for momentum mapping of the full half-space, based on a large single hemispherical analyzer (path radius 225 mm). Excitation by an unfocused He lamp yielded an energy resolution of 7.7 meV. The performance is demonstrated by k-imaging of quantum-well states in Au and Xe multilayers. The $\alpha^2$-*aberration term* ($\alpha$: entrance angle in the dispersive plane) and the *transit-time spread* of the electrons in the spherical field are studied in a large pass-energy (6 to 660 eV) and angular range ($\alpha$ up to $\pm 7°$). It is discussed how the method circumvents the preconditions of previous theoretical work on the resolution limitation due to the $\alpha^2$-term and the transit-time spread, being detrimental for time-resolved experiments. Thanks to k-resolved detection, both effects can be corrected numerically. We introduce a *dispersive-plus-ToF* hybrid mode of operation, with an imaging ToF analyzer behind the exit slit of the hemisphere. This instrument captures 3D data arrays $I(E_B,k_x,k_y)$, yielding a gain up to $N^2$ in recording efficiency (N being the number of resolved time slices). A key application will be ARPES at sources with high pulse rates like Synchrotrons with 500 MHz time structure.

## I. INTRODUCTION

During the last decade, photoelectron momentum microscopy (MM) with energy dispersion by double-hemispherical analyzers or time-of-flight (ToF) recording has proven to be a powerful approach for angle-resolved photoelectron spectroscopy (ARPES). In this special type of low-energy electron microscope, momentum distributions are observed directly in k-space. MMs record the high-resolution Fourier image (spanned by the two transversal momentum components $k_x$ and $k_y$) in the backfocal plane of a special objective lens. Various types of objective lenses have been optimised for certain parameters, targeting different science cases. High-performance ARPES requires high k-resolution in the Fourier image, the best value for a MM reported by Tusche et al. is 0.005 Å$^{-1}$ [1]. Large k-fields of view exceeding diameters of 16 Å$^{-1}$ and high kinetic energies up to > 7 keV have been achieved with special high-energy objectives [2]. Both together are required for hard-X-ray photoelectron diffraction, where MM has reached a new level of recording speed [3]. Enlarged k-fields and high energies are also necessary for efficient bulk band mapping, eliminating the strong diffraction modulation of the valence-band patterns [4]. Further design criteria aim at a field-free sample region and maximum working distance (e.g. 27 mm field-free space in [5]), which is advantageous for non-planar samples like cleaved microcrystals or other 3D structures.

Alongside with the different types of objective lenses, there is a fundamental technical difference concerning the method of energy discrimination. One type employs a dispersive energy filter consisting of a double-hemispherical analyser (double-HSA), first realized by Kroemker et al. [6] using the NanoESCA, originally developed for spectroscopic real-space imaging [7,8] exploiting chemical information from core levels. This basic approach has led to a new generation of a high-resolution momentum microscope at the Max-Planck Institute in Halle (Germany) [1,9].

The double-HSA instruments produced a series of benchmarking results on the surface electronic structure of high quality single crystals [1,10], topological insulators [11-13], layered semiconductors [14], and strongly correlated materials [15]. The principle of momentum microscopy, by virtue of its simultaneous imaging of all photoemission directions has proven very useful for the quantitative analysis of photoemission intensities and their comparison with theoretical models [16-18], as well as adsorbate interactions and dichroism effects [19,20]. The rapid recording of whole momentum images allows the study of reactive or short-lived surfaces with high momentum and energy resolution [21], while the real-space imaging mode gives access to heterogeneous systems [11].

The second type of MM is based on time-of-flight (ToF) detection, facilitating parallel recording of a certain energy band. The ToF-MM solution emerged from earlier work on ToF-PEEM (photoemission electron microscopy) at the University of Mainz (Germany) [22-24]. Naturally, the ToF-approach requires sources with defined time structure, whereas the hemisphere instruments can be operated with cw-sources like discharge lamps. High-performance photon sources like Synchrotrons, free-electron-lasers (FELs) or high-harmonic-generation (HHG) lab sources are characterized by a precise time structure. At sources with repetition rates > 100 kHz, the ToF-approach bears the advantage of high data throughput due to 3-dimensional recording. The full 4D spectral density function and Fermi-velocity distribution, comprising 20 data arrays with ~$10^6$ ($k_x$,$k_y$,$E_B$)-voxels each can be measured in short time at a high-brilliance soft-X-ray beamline, as demonstrated by Medjanik et al. [25]. ToF-MM has been used in a large spectral range, giving access to surface states excited with vacuum UV light [26-29], plasmonic emission from nanoobjects [30], circular dichroism texture in the soft-X-ray range [31] and detail-rich full-field XPD diffractograms in the hard X-ray range [3,32,33], opening the door to use a MM for true bulk band mapping [4,34,35].

A key advantage of ToF recording at pulsed sources is the capability of observing ultrafast processes. Very recently, ToF-MM with fs-pulses from the FEL FLASH at DESY, Hamburg, was successful in observing ultrafast electron dynamics [36,37] including core-levels [38], ultrafast orbital mapping [39] and time-dependent XPD [40] in a single setup. Experiments with the same type of instrument using femtosecond HHG-sources in several groups yielded first impressive results on time-resolved photoemission [41-45] and molecular orbital mapping [46]. For this new generation of multidimensional photoemission experiments data post-processing is an important issue [47].

A particular stronghold of MM is the combination with imaging spin-polarimetry. Suga and Tusche give an overview of the increase of the effective figure-of-merit of imaging spin filters (see Fig. 28 in [48]). After initial work with W(001) spin filters [49,50], the advent of Ir(001) [51,52] and Au-monolayer-passivated Ir(001) [1,53,54] as spin filter surfaces has led to efficient devices for spin-texture analysis. Impressive results have been obtained with double-HSA type instruments using lab sources [1,13,55-57] and Synchrotron radiation at ELETTRA, Trieste [58-61]. Spin-resolved band mapping using the ToF-MM approach has been performed in the UV using fs-laser radiation [62], in the VUV [63-67] at BESSY, Berlin, in the soft-X-ray range [68] and very recently using hard X-rays [69] at PETRA III, Hamburg.

The common belief that k-microscopy with dispersive analyzers requires a double-HSA tandem configuration in order to compensate the sphere aberrations [6-8] has been questioned by Tusche et al. [9]. Their analytical treatment reveals that the refraction of the electron rays due to the fringe fields counteracts the aberrations and compensates them up to $5^{th}$-order terms. In turn, the sphere aberrations are negligible and the second hemisphere can rather be used to improve the dispersion as demonstrated experimentally [9]. An important consequence of these findings is the fact that a single hemisphere should be suitable for momentum imaging as well. A single-HSA with twice the radius as the two HSAs in a tandem configuration, thus, can achieve a similar dispersion. In addition, the electron optics is simpler because the transfer lens between the two hemispheres is not needed.

Here we show the first MM results taken with a large hemispherical analyzer with 225 mm path radius [70]. It is demonstrated how the action of the leading aberration term (the so-called $\alpha^2$-term, $\alpha$ being the entrance angle in the dispersive plane of the analyzer) is naturally resolved and can be tuned via the angular filling of the analyzer. The $\alpha^2$-term can either be reduced until it is negligible or alternatively be eliminated numerically. The angular beam spread and lateral magnification are reciprocal to each other (Helmholtz-Lagrange invariant); for details, see [1]. Hence, large angular filling is advantageous for large photon-beam footprints on the sample surface.

Finally, we introduce a new *dispersive-plus-ToF* hybrid type of momentum microscope, combining the advantages of both instrument families. A ToF analyzer is implemented behind the exit slit of the hemisphere. Extensive work over decades ([71-79] and references therein) was devoted to the detrimental action of the *transit-time spread* in HSAs, caused by the different path lengths on Kepler ellipses at finite angular acceptance. In our solution, we circumvent the precondition of this work by imaging the Fourier plane instead of unifying all these different rays on the detector in the exit plane. In turn, the time spread is resolved and can be precisely eliminated. The transit time is strictly deterministic and to a good approximation a linear function of the $k_y$-component. This hybrid mode is favourable at photon sources with pulse rates, which are too high for pure ToF recording. Very common sources of this type are Synchrotrons operated at a pulse rate of 500 MHz. In connection with a new delay-line detector with improved time resolution, the *dispersive-plus-ToF* hybrid mode improves the recording efficiency (resolution and / or transmission) by 1-2 orders of magnitude.

## II. TECHNICAL BACKGROUND

**II.A. Interplay of $\alpha^2$-aberration, resolution and phase-space acceptance**

The field of electron spectroscopy is large dominated by the use of hemispherical analyzers, because they have a number of obvious advantages in comparison with other types of dispersive spectrometers. The solution of the equation of motion in the 1/r potential leads to Kepler ellipses, the properties of which are perfectly understood. Motion in an ideal spherical field is characterized by first order focussing after 180°. Hence, placing entrance and exit slits under 180° and correcting the fringe fields by suitable electrodes (see next section) yields an energy analyzer with excellent performance. Today, the exit slit is mostly replaced by a 2D imaging detector, allowing to record a certain energy band (along the dispersive direction) and angular range (in the non-dispersive direction) in parallel. All details of the HSA have been studied theoretically over many decades, see e.g. [71-79,9] and references therein. An excellent introduction into the principle of this analyzer can be found in [80].

The leading aberration term ($\alpha^2$-*aberration*) for charged particles moving in a HSA reflects a property of these Kepler orbits. Electrons with pass energy $E_{pass}$ entering the analyzer at an entrance angle $\alpha = 0$ travel on a circular path with radius $R_0$ and reach the exit plane after 180° at radius $R_\pi = R_0$ (ray 1 in Fig. 1). However, electrons with the same energy entering with $\alpha \neq 0$ run on ellipses and fall short ($R_\pi < R_0$) of the exit slit (rays 2 and 3). First-order focussing ensures that the term linear in $\alpha$ vanishes after 180°. Here $\alpha$ denotes the entrance angle (measured in the dispersive plane) after refraction of the electron rays at the entrance aperture. The angular dependence of the deviation from $R_0$ is given by eq. (1), where the prefactor K depends on details of the fringe-field correction [72,78,9]:

$$R_\pi(\alpha) = R_0 - K \cos^2\alpha \qquad (1)$$

The $\alpha^2$-aberration is a major resolution-limiting factor for HSAs in the conventional mode of operation (without resolving $k_\parallel$). At given values of $R_0$ and $E_{pass}$ the energy resolution $\delta E$ depends not only on the entrance and exit slit widths $W_1$ and $W_2$ but also on the square of the half angle $\alpha_{max}$ of the ray bundle entering the analyzer. The $\alpha_{max}^2$-*term* appears as second term in the energy resolution formula [71]:

$$\delta E_{FWHM} = E_{pass} [(W_1+W_2)/4R_0 + \alpha_{max}^2/2] \qquad (2)$$

According to the "Kuyatt-Simpson criterion", the $\alpha_{max}^2$-term should not exceed 0.5 times the "slit"-term. Hence, in a conventional HSA high resolution requires to limit $\alpha_{max}$ typically to 1.5 to 2.5°, depending on the desired resolution. This restriction of the opening angle is important for the phase-space acceptance. As discussed in detail in [9], the $\alpha_{max}^2$-term does not apply for a MM because $\alpha$ is resolved in the k-image. We notice that the present 225 mm analyser [70] has reached a resolution of 1.6 meV in conventional mode of operation [81].

For the second angular coordinate $\beta$ the situation is simpler because all electrons deviating from the path $R_0$ by an angle $\beta$ in the plane perpendicular to the dispersive plane (see coordinate system in top right inset of Fig. 1) travel on equatorial circles (if $\alpha = 0$) or on identical ellipses (if $\alpha \neq 0$). Hence, $\beta$ does not enter the resolution equation. In their detailed treatment of aberration coefficients, Wannberg et al. [72] discuss that some higher-order aberration terms are connected with the angle $\beta$ as well. In order to avoid confusion, we do not introduce the emission angle at the sample surface in Fig. 1 because in a momentum microscope the objective lens converts the emission angle into the transversal momentum components $k_x$ and $k_y$.

Besides the effect of $R_\pi < R_0$ for $\alpha \neq 0$ and the $\alpha_{max}^2$-term in the resolution equation there is a third manifestation of the properties of Kepler ellipses. The exit angle $\alpha_{exit}$ behind the analyzer is not identical to the entrance angle $\alpha$ but depends also on the exact exit position $R_\pi$, which is a complex function of starting angle and kinetic energy. In k-microscopy, the image information is encoded in terms of angular coordinates. Hence, the complex interplay of $\alpha_{exit}$, exit position $R_\pi$ and kinetic energy can cause an aberration in the momentum image and might set a limit to the k-resolution. This special type of image aberration is termed *sphere aberration* and has been discussed by Tonner [76].

Fig. 1 shows a schematic view of the single-hemisphere momentum microscope with entrance and exit optics and schematic rays in analyzer and lens optics. We use the identical lens geometry as described in detail in [1]. It comprises an objective lens (cathode-lens type with extractor field) and two lens groups adjusting beam energy, image size and angular magnification in the entrance plane. Electrons with energy $E_0$ passing the entrance aperture perpendicular to the slit plane travel on a circular trajectory (path 1) with radius $R_0$ and pass the exit aperture. Electrons with the same energy $E_0$ on paths deviating from this central ray by an angle $\alpha$ (measured in the dispersive plane y-z) travel on Kepler ellipses. These electrons fall short and cannot pass the exit slit, no matter whether they travel closer to the inner (path 2) or outer hemisphere (path 3). They pass the exit slit when the analyzer voltages are set to a nominally lower energy $E_{pass}= E_0-\Delta E$. This effect is referred to as *non-isochromaticity* and is a serious aberration in conventional real-space imaging spectrometers [82]. The energy shift is a parabolic function of $\alpha$ and can thus be corrected numerically in the $\alpha$-resolving detection mode of a k-microscope as shown in [9].

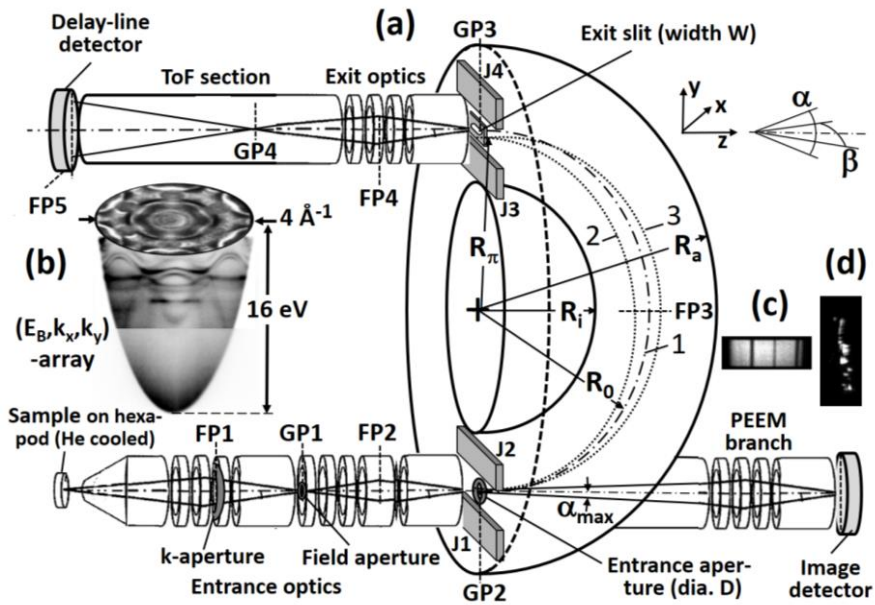

**FIG. 1.** Schematic view illustrating the basic principle of the single-hemisphere momentum microscope. (a) Set-up comprising entrance zoom optics, straight branch for real-space imaging (PEEM), hemisphere and exit optics with time-of-flight discrimination behind the exit slit. J1-J4 denote Jost plates for fringe-field correction. Electrons entering the analyzer at $\alpha= 0$ travel on a circular path (ray 1) with radius $R_0$. The effect of the $\alpha^2$-aberration (rays 2 and 3 fall too short) is exaggerated. $\alpha$ and $\beta$ (top right inset) denote the entrance angles with respect to the central ray in the dispersive plane y-z ($\alpha$) and perpendicular to this plane x-z ($\beta$). GP1-GP4 denote the Gaussian planes (real-space image planes), FP1-FP5 the Fourier planes (momentum-image planes). (b) Data array recorded with He I radiation, (c) PEEM image of entrance aperture with calibration-grid lines in plane of the field aperture and (d) hot-spot emission along a scratch on a smooth single-crystal surface (excitation by ps-laser).

Aberrations and non-isochromaticity pose limits not only on energy and momentum resolution but also on the accessible phase-space interval. The relevant phase-space is defined by the generalized coordinates of electron velocity (proportional to $\sqrt{E}$), opening angle (proportional to $\sin \alpha$) and lateral magnification (M). The *Helmholtz-Lagrange invariant* of paraxial optics is given by the product of refraction index, aperture angle and lateral size, translated into particle optics as:

$$M \sin\alpha_{max} \sqrt{E} = \text{constant} \qquad (3).$$

The desired resolution dictates pass energy and size of the analyzer slits. Hence, the accepted phase-space volume can be optimized by choosing magnification M and filling angle $\alpha_{max}$. In addition, the total recording efficiency depends crucially on the fraction of emitted photoelectrons that actually pass the entrance aperture. For photon-beam diameters in the range of 100 µm or more, the image of the photon footprint in the analyzer entrance plane is larger than the entrance aperture so that part of the signal is cut. In this scenario, small lateral magnifications M are desirable, demanding large filling angles $\alpha_{max}$. A central question of the present work is to elucidate up to which angles $\alpha_{max}$ good k-imaging is possible.

In summary, the properties of Kepler ellipses manifest in various ways in conventional and k-resolving HSAs. The $\alpha^2$-aberration leads to $R_\pi(\alpha\neq 0) < R_0$ (eq. (1)), the $\alpha_{max}^2$-term limits the resolution when $\alpha$ is not resolved (eq. (2)), the sphere aberration affects $\alpha_{exit}$ and might distort angle-encoded k-images and the non-isochromaticity leads to a variation of the kinetic energy across an angular-encoded momentum or real-space image.

## II.B. Fringe-field corrections, refraction of electron rays at analyzer entrance and exit

In an ideal spherical field, perfect first order focussing is achieved after 180°. In reality, however, the spherical field is terminated by electrodes carrying the entrance and exit slits or the detector. A tremendous amount of work has been devoted to the action of the fringe fields, launched by the early work of Herzog [83,84]. Uncompensated fringe fields induce a focussing effect and shift the focal point away from 180° [74,84-87].

Jost [88,89] modelled the fringe fields by a resistor network and proposed a geometry which brings the focus back to 180°. Zouros et al. [77,90-95] performed detailed studies of the equation of motion, with particular emphasis on the various approaches of fringe-field correction and the temporal spread (cf. next section). They propose an interesting paracentric geometry, which unfortunately is not suitable for k-microscopy because it introduces additional astigmatism. Hu and Leung refined the Jost design of fringe-field correction with emphasis on the use of position-sensitive detectors [87]. Tusche et al. [9] revisited the issue of the fringe-field effect and considered the refraction of the electron rays at the entrance and exit apertures. Here we recall the essential facts, which are relevant for the present instrument.

For a momentum microscope, the fringe-field correction is crucial for optimization of both energy and k-resolution. The entrance optics of a MM focuses the beam such that there is a Gaussian (real-space) image of the sample surface / photon footprint in the entrance aperture. Further, there is a reciprocal image (k-image) exactly in the centre of the optical path in the analyzer (Fourier plane FP3 in Fig. 1). This assures a one-to-one transfer of the angular-encoded 2D momentum image from the entrance to the exit aperture. Since the spherical field contributes to the focussing, the fringe-field correction is very important for this delicate interplay of beam shaping by the entrance optics and quality of the k-image on the delay-line detector at end of the ToF section (FP5 in Fig. 1). Notice that for k-imaging real and reciprocal planes are interchanged; see, e.g., Fig. 1 in the paper by Tonner [76], which describes energy-filtered real-space imaging.

The so-called *Jost-plates* are designed such that the field in the fringing region is enhanced in order to compensate the field termination due to the entrance and exit "short cuts". Modelling the fringe-field behaviour, Jost arrived at a geometry (Fig. 5 in [88]), where the *integral* field action along the central ray path is identical to that in an ideal spherical field. The HSA in our set-up [70] uses the geometry proposed by Jost, but the potentials of the four plates (J1-J4 in Fig. 1) are treated as free parameters for optimization. As discussed in the early work by Wannberg et al. [72,78], this degree of freedom is necessary to optimise the fringe-field compensation. The Jost design, in particular, strictly holds for the homogeneous field of a parallel plate capacitor and a rectangular system of coordinates, instead of the 1/r potential and spherical symmetry of the beam axis. Indeed, we find a substantial deviation in the optimum values from the original design (plate potentials equal inner- and outer-sphere potential). The potential of the outer plates is more than 20% away from that of the outer sphere. In addition, the potential settings of J1 at entrance and J4 at the exit (Fig. 1) are systematically different.

Inclusion of the refraction of the electron rays in the entrance and exit slit in [9] revealed a previously overlooked relation between the change of the ray paths due to refraction and the leading term of the image aberrations. The authors arrive at the result that the refraction counteracts the sphere aberration, which in the past has been considered as the most severe source of image distortions in hemispherical energy filters [76]. In the framework of an analytical treatment, both contributions even compensate each other for angles up to $\alpha_{max} \approx 3°$. The analytical expressions further reveal that for both in-plane image coordinates the higher-order aberration terms are negligible. In particular, the next aberration term, proportional to $\alpha^4$, is less than $10^{-3}$ for angles up to 10°. This is smaller than the experimental k-resolution. The compensation of sphere aberration and refraction falsifies our earlier paradigmatic claim [7,8] that an inverting lens and second hemisphere are required for aberration correction.

Moreover, the non-isochromaticity can be easily corrected numerically [9], since it is a simple expression depending just on the square of the momentum component $k_y$. For large slit widths above ~1 % of the analyzer radius, the $\alpha^2$-term is negligible compared to the energy resolution defined by the slits.

**II.C. Transit-time spread of electrons on different Kepler orbits**

The time spread of electrons passing a HSA has been studied theoretically in much detail (see pioneering work of Imhof et al. [73] and later extensive work of Zouros et al. [77,79] with references therein). For the electrons travelling on Kepler orbits the transit time between entrance and exit slit depends on the entrance angle $\alpha$ with respect to the central ray. Electrons passing the entrance slit perpendicular to the slit plane ($\alpha = 0$) travel on a circular orbit with radius $R_0$ and constant energy $E_{pass}$. These electrons reach the exit slit after the transit time $t_0 = R_0\pi (2E_{pass}/m)^{-1/2}$, m being the electron mass. For the given analyzer ($R_0$= 225 mm) we have $t_0$= 238 ns at $E_{pass}$= 100 eV. Electrons entering with $\alpha > 0$ (ray closer to the outer hemisphere) travel a longer path and the mean energy is smaller, leading to larger transit time. For $\alpha < 0$ the path is shorter, the energy higher and the time shorter. In addition, finite slit widths D lead to different radii $R_0$ and hence to different transit time. Imhof et al. [73] give an approximate equation for the full width at half maximum of the *time-spread profile* as

$$\Delta t_{FWHM} / t_0 \approx 0.60 \, (D/R_0) + 2.23 \, (1 - 2.79 \, D/R_0) \, \alpha_{max} \qquad (4).$$

For our analyzer, the first term is 0.5 % and the second 3.9 % for the largest slit of 2 mm and $\alpha_{max}$= 1°, which together yields $\Delta t_{FWHM}$= 10.5 ns for $E_{pass}$= 100 eV. This value reduces to 4.1 ns for $E_{pass}$= 660 eV, a typical pass energy for the ToF hybrid mode.

For the correction of the time spread, it is important to know the analytical dependence of the transit time $t_\pi$ as function of the angle $\alpha$. Sise and Zouros give a power expansion of the transit time as function of $\alpha$ for the general case of the paraxial entry condition [79]. Here we show their result for the present instrument (in their terminology, the conventional hemispherical analyzer). In the equation for the *transit time $t_\pi$*, we drop the terms resulting from the slit widths because these are negligible for practical work with slits of 200–500 μm.

$$t_\pi(\alpha) \approx t_0 [1 + 4(\alpha/\pi) - 10\alpha^3/3\pi] \qquad (5).$$

This equation predicts an essentially linear dependence of $t_\pi$ on the entrance angle $\alpha$ in the dispersive plane in the relevant angular range. The change of transit time is 2.2 % for $\alpha$= 1°; for $E_{pass}$= 50 and 660 eV we expect changes of 7.4 and 2.0 ns. The third-order correction is negligible at our working conditions; it reaches 0.5 % for $\alpha = 10°$.

Increasing linearly with path radius and with $E_{pass}^{-1/2}$ the transit-time spread reaches tens of nanoseconds at high-resolution conditions. This time spread is irrelevant in normal spectroscopy, but prohibitive for experiments demanding high time resolution in electron detection, e.g. in time-resolved photoemission (pump-probe experiments) or in coincidence experiments. The problematic time spread can be somewhat reduced by using paracentric entry conditions [77,79], where the incoming ray bundle is displaced from the conventional ideal entry. Kroemker et al. [6] have shown that a tandem configuration of two hemispherical analyzers with an inverting lens in between can reduce the time spread by an order of magnitude in comparison with a single HSA with same radius. For an analyzer with radius 124 mm, $E_{pass}$= 100 eV and $\alpha$= 1.5° ref. [6] predicts a time spread of 500 ps. Volkel and Sander [75] propose a tilted orientation of the detector, in order to cope with the tilted (and curved) isochrone surfaces. Here we show that time-resolved detection in the Fourier plane opens the door to a perfect correction of the transit-time spread, with the precision only limited by the time-resolution of the detector (< 100 ps).

### III. SETUP AND CHARACTERIZATION

**III.A. Layout of the single-hemisphere momentum microscope, details of recording scheme**

The present instrument results from development at three research institutes in cooperation with two company partners. Work at the Max-Planck Institute in Halle (Germany) focused on the electron-optical column in combination with a double-HSA and a laboratory photon source; a benchmark was set in terms of k-resolution and spin-filter efficiency [1]. This development is continued at Forschungszentrum Jülich (Germany), in particular the fundamental work on the interplay of refraction at the entrance and exit slit and sphere aberration [9]. At Mainz University the ToF-MM technique has been established, exploiting the time structure of Synchrotron radiation in various spectral ranges [25,26,3] and at the FEL source FLASH [36]. The core expertise of the company partner Surface Concept GmbH Mainz concerns delay-line detector development [23,96] and SPECS GmbH Berlin develops novel concepts for HSAs [5,70] and fine-focused laboratory sources [97].

The entrance optics (schematic sketch see Fig. 1) is identical to the Halle instrument; details are given in [1]. The objective lens forms an achromatic reciprocal image in Fourier plane FP1, where a set of adjustable k-apertures serves for definition of the k-field of view and for optimisation of contrast and resolution in real-space imaging mode (PEEM mode [98]). An auxiliary grid serves to focus the subsequent optics exactly to the correct plane. Apertures and grid are adjusted by a pair of linear piezomotors. An intermediate lens group forms a real-space image of the sample surface in the first Gaussian plane GP1. This plane hosts a set of nine piezomotor-driven field apertures, allowing precise definition of size and position of the region of interest (ROI) on the sample surface. The sizes of the field apertures correspond to ROIs between ~70 μm down to the μm-range on the sample surface. According to the calculations in [1] this sophisticated entrance lens system can achieve ROIs in the k-imaging mode in the range < 1 μm for high extraction fields and moderate k-field diameter (see Fig. 2 in [1]). Using the identical lens system, small ROIs in the few-μm-range have been reached in practice with sufficient photon density. The present measurements have been carried out with a moderate extractor field (~2 kV/mm). For ultimate k-resolution the extractor field can be increased to 8 kV/mm. We notice that a modified type of entrance lens can also work with zero field in the sample region [2], at the expense of accessible angular interval.

Like in the backfocal plane, an auxiliary grid in GP1 facilitates precise adjustment of the downstream optics to this plane. A third lens group (five-element zoom lens) focusses a real-space image of sample surface and field aperture (or grid in plane GP1) to plane GP2, coinciding with the entrance plane of the analyzer. This zoom lens can be adjusted in a wide range of lateral and angular magnifications and adapts the beam energy to the pass energy of the analyzer. The instrument comprises a large single HSA with a mean path radius of $R_0$ = 225 mm, whereas the Halle system employs a double-HSA with $R_0$ = 150 mm each. The present analyzer has a set of eight apertures with diameters between 200 μm and 2 mm in the entrance and corresponding slits with widths between 200 μm and 2 mm in the exit, coupled by a rotatable selector mechanism.

The upper branch behind the exit slit comprises the exit optics (five-element zoom lens), a low energy drift section for ToF dispersion and a time-resolving image detector (delay-line detector DLD). Most measurements have been made with a normal DLD, but we also worked with the first prototype of a new design with improved time resolution (preliminary value 70 ps [96]).

The sample is mounted on a motorized precision hexapod goniometer with liq.-He cooling ($T_{min}$ < 10 K); the resolution measurements in Fig. 2 have been performed at 14 and 22 K. Base vacuum was $10^{-10}$ mbar. Photoelectrons are excited by VUV-photons from an unfocussed capillary discharge source with a glass capillary ending about 35 mm from the sample. Here we show data for He I (21.23 eV) radiation, but He II (40.81 eV), Ne I (16.85/16.67 eV) and Ar I (11.83/11.62 eV) were also used. From the divergence of the photon beam measured for this type of source (Fig. 4 in [99]) and the impact angle of 68° off normal we estimate a large photon spot size of 2 x 0.8 mm$^2$ on the sample surface.

The straight branch is used for real-space imaging (PEEM) without energy discrimination, with the analyzer voltages switched off. Here we use a simple multichannelplate–screen–CCD camera combination as image detector. The PEEM-mode is important for inspection of the sample surface, in particular for inhomogeneous or structured samples. Further, it is ideal for checking position and size of the photon beam on the sample. Finally, it serves to fulfil the basic focusing condition of the instrument, namely to bring into coincidence the Gaussian images of (i) the sample surface, (ii) the field aperture in the first intermediate image plane and (iii) the entrance aperture of the analyzer. The inset Fig. 1(c) shows the coincidence of the images of the auxiliary grid in plane GP1 with the entrance and exit slits. This image was taken using the detector behind the exit lens system and drift tube. Its outer shape is rectangular because the exit aperture is a slit (upper and lower edges are the rims of the exit slit) and the continuous energy band appears as vertical stripes, the width given by the entrance aperture (left and right edges are the rims of the stripes). For the same reason only the vertical division bars of grid meshes are visible. In this mode, we can set a smaller virtual exit slit inside of the true exit slit via the DLD software in k-integrated recording.

The grid in the plane of the field apertures (vertical stripes in Fig. 1(c)) serves as calibration standard for setting the desired value of the lateral magnification. Given the spacing of the grid bars of 254 μm (mesh 100) and the selected exit slit width of 500 μm we arrive at a magnification of 2 between image planes GP1 (field aperture plane) and GP2/GP3 (slit planes) in the example of Fig. 1(c). The field aperture has the function of a contrast aperture for the k-image and thus defines the k-resolution. The analyzer slits define the energy resolution at a given pass energy. The lateral magnification is inversely proportional to the angular magnification, which defines the non-isochromaticity (section III.C). All three parameters must be controlled and precisely set to the proper values. The independent selection of field aperture, magnification and analyzer slits is particularly important for the dispersive-plus-ToF hybrid mode.

The cathode lens converts the photoelectron angular distribution into a circular $k_\parallel$-image with maximum usable diameter of ~8 Å$^{-1}$ for the present entrance optics. At kinetic energies up to 70 eV the k-space interval of the present optics corresponds to the full half-space in front of the sample surface. At a typical soft X-ray energy of 500 eV the k-field of radius 4 Å$^{-1}$ corresponds to a solid-angle interval with 21° polar angle and full 2π azimuth. Notice that a different lens system can record larger k-fields [2,3]. The entrance lens forms a magnified image in the entrance aperture (diameter D) of the analyzer. Thus, the sample region contributing to the signal has a diameter of D/M, where M denotes the total magnification of the entrance optics.

Due to the $\alpha^2$-term, all electrons on paths with $\alpha \neq 0$ fall too short and thus appear in the recorded patterns like slower electrons. This quadratic $\alpha$-dependence of the non-isochromaticity is directly observed in $k_\parallel$-resolved recording [9]. The constancy of the product $M \sin\alpha$ at a given pass energy (eq. (3)) offers the possibility to optimise the recording efficiency. There is no "ideal" set of parameters, because the interplay of recording efficiency, magnification and angular filling can be optimised only under the boundary conditions of the desired k-field-of-view and the effective source area on the sample surface (ROI). The latter is defined either by the footprint of the photon beam, or by the size of the field aperture or by the size of the entrance aperture of the analyzer, depending on which of the three is the limiting one.

The angular spread of the ray bundle in the analyzer entrance can be tuned in a wide range, obeying eq. (3). We have studied this interplay of lateral magnification and angular filling for many different pass energies (from 6 eV to 660 eV) and slit widths (from 200 μm to 2 mm); their combination defines the energy resolution. The goal was to elucidate limitations of parameter space, with special emphasis on a possible upper limit for the angular filling. The *maximum filling angle* $\alpha_{max}$ of the ray bundle in the entrance slit of the analyzer is given by

$$\alpha_{max} = \arcsin [\, 1.95 \, k_{\parallel,max} \, E_{pass}^{-1/2} \, M^{-1} \,] \qquad (6)$$

Here $k_{\parallel,max}$ denotes the radius of the recorded k-field-of-view (in Å$^{-1}$), $E_{pass}$ the pass energy (in eV) and M the lateral magnification of the Gaussian image in the entrance plane of the analyzer with respect to sample coordinates. This relation derives directly from the Helmholtz-Lagrange invariant (eq. (3)).

Please notice that on the sample side only $k_{\parallel,max}$ enters this equation. For the operating conditions of the analyzer it is irrelevant, whether the momentum microscope records the photoelectrons emitted into the full half space using He I excitation ($E_{kin} \approx 16$ eV), or into a cone of +/-10° at a typical soft-X-ray energy like $E_{kin}$= 500 eV or into a cone of +/- 3° at a hard-X-ray energy of $E_{kin}$= 7000 eV. The latter value is the highest energy we recorded using a high-energy version of the ToF-MM [3]. Given the selected pass energy and magnification, the analyzer "sees" only the value of $k_{\parallel,max}$, being 2 Å$^{-1}$ in all three cases. Moreover, all but two potentials of the instrument are kept fixed, even when changing $E_{kin}$ in such a large range. If the sample is biased (positive) at $U_{sample}$= $E_{kin}$/e the potential of the analyzer slits is kept fixed at $U_{slit}$= $E_{pass}$/e. Then only the focus of the first lens (objective) has to be adjusted when the energy varies strongly. For sweeps in small energy intervals of several eV, like in the present work, the focus voltage can be kept constant. At sufficiently high extractor potential, the k-image in the backfocal plane is achromatic in a range of several 10 eV. In turn, during energy scans only a single voltage, the sample bias, needs to be swept.

In the practical application of eq. (6), first the desired k-field of view is selected via the size of the k-aperture in the 1$^{st}$ Fourier pane FP1 (Fig. 1); this defines field radius $k_{\parallel,max}$. This aperture does not need to be centered in the $k_\parallel$ image ($\bar{\Gamma}$-point), but it can also be placed off-centre around, e.g., a $\bar{K}$- or $\bar{M}$-point (cf. Figs. 6,7) or the $\bar{\Gamma}$-point of the second Brillouin zone. Next, $E_{pass}$ is chosen, defining the resolution together with the entrance and exit apertures. Finally, the filling angle $\alpha_{max}$ of the analyzer is set using the zoom-optics of the entrance lens, controlled by the size of the final k-image with the exit optics switched off. According to eq. (6), $\alpha_{max}$ defines the magnification M. A small $\alpha_{max}$ is favourable for small non-isochromaticity (cf. Section III.C.). However, large $\alpha_{max}$ is in favour of high intensity if the photon footprint at the sample surface is large.

At state-of-the-art Synchrotron beamlines and laser sources, the size of the photon footprint is of the order of 10 to 50 μm. Then the full photoelectron signal from the illuminated region is captured with magnifications up to M ~ 10. When the photon footprint is large, like in the present experiments, the entrance aperture blocks part of the Gaussian image. In this case M should be as small as possible in order to detect the maximum fraction of the total photoelectron signal. Then eq. (3) emphasizes to use large values of $\alpha_{max}$. Given the circular entrance aperture, the fraction of the total photoelectron signal entering the analyzer is proportional to $M^{-2}$. The unfocussed capillary discharge lamp used for the present study is only a test source for later synchrotron excitation. For most settings, the detectable signal when using this lamp is only a small fraction of less than 1% of the total photoelectron yield. At the lowest magnification studied (M= 3) the smallest entrance aperture (200 μm) translates to a ROI of ~70 μm dia. on the sample surface, in our setup a fraction of only 2.5 x $10^{-3}$ of the illuminated area (photon footprint ~1.6 $mm^2$).

Assuming a homogeneous illumination of the entrance aperture (diameters D= 200 μm to 2 mm), the electron intensity entering the analyzer depends quadratically on its diameter (area $D^2 \pi/4$). The exit slit (widths W= 200 μm to 2 mm) selects the bandwidth, hence, the electron intensity behind the slit depends linearly on the slit width W. In the unfavourable case of a large photon footprint, a factor of 2 reduction in the sizes of D and W costs a factor of 8 in intensity, which we confirmed experimentally throughout the entire pass energy range studied. In the opposite limit, where the Gaussian image of the photon footprint lies completely inside of entrance aperture D, the intensity loss factor is only 2. In practical work at state-of-the-art Synchrotron beamlines in the soft X-ray range [100,101] or when using a well-focussed laboratory source, the intensity loss will lie between the extremal cases 8 and 2.

Increasing the resolution via subsequent ToF-analysis avoids this reduction of intensity. Instead of increasing resolution by closing the apertures, the subsequent ToF analyzer ("ToF-booster") disperses the electron distribution passing the exit slit. A key application of the dispersive-plus-ToF hybrid mode is ARPES at synchrotrons with 500 MHz filling pattern, i.e. 2 ns gap between adjacent photon pulses. Given the time resolution of the new generation of DLDs (~70 ps), the ToF analyzer has a nominal resolution of 28 energy slices. In practice 20-25 of these are usable, in order to avoid overlap at the borders of the time gap. Quantitative examples are given in Section III.D.

### III.B. Energy and momentum resolution

3D ($E_B$,$k_x$,$k_y$) band-structure patterns in the half-space above the sample surface are recorded in terms of many full-field images in energy steps covering the desired energy interval. Energy increments are chosen sufficiently small (1-2 meV for high-resolution scans, 10-20 meV for wide energy ranges), so that the 3D stacks appear quasi-continuous. Nominal resolution values cover a large range from $\delta E$= 2.7 meV at $E_{pass}$= 6 eV and 200 μm slit to $\delta E$= 2.9 eV at $E_{pass}$= 660 eV and 2 mm slit (cf. Eq. (2)). Values of $\delta E$ in the eV-range are of interest for the dispersive-plus-ToF hybrid mode of operation, Section III.D.

Figure 2 shows a sequence of energy-resolution measurements at pass energies between 100 eV and 6 eV. For pass energies of 100, 50 and 25 eV (a,b,c, respectively) and smallest aperture of 200 μm the measured resolution agrees with the nominal resolution derived from eq. (2). For measurements with $E_{pass}$= 6 eV and apertures of 2, 1 and 0.5 mm (d,e,f, respectively) thermal broadening (4 meV at 14 K) is significant and hence the measured width of the Fermi-energy cutoff is somewhat larger than expected from eq. (2). The intrinsic line width of the He I line is estimated to be 2 meV [102]. For these measurements the discharge lamp was operated at reduced current and rather high pressure (but before self-absorption sets in). Assuming that all contributions add up as a sum of squares, we obtain 7.7 ± 1 meV for the analyzer resolution limit in (f). This value compares reasonably well with the expected resolution of 6.7 meV according to eq. (2). Due to the unfocussed light source the signal-to-noise ratio in the measurements at 6 eV pass energy was rather low and thus it was not possible to use slits smaller than 500 μm. For higher photon-flux density, e.g. from a focussed He lamp or at a Synchrotron beamline, we expect an improvement of resolution by more than a factor of two for 200 μm apertures. Hence a resolution of few meV seems realistic for smallest aperture and pass energies that still allow good k-imaging (examples in Fig. 3). For pulsed narrow-band sources, where the ToF-booster can be used, the resolution limit can be reached at larger pass energies and larger slits because the transmitted energy band is cut into time slices (cf. Section III.D).

The k-resolution of the instrument has been determined by line profiles across the rim of the photoemission horizon, yielding a width of 0.008 Å$^{-1}$. The Shockley surface state on the evaporated Au films (cf. Section IV) is not suitable for resolution determination, since it is broadened due to the small size of the domains in the film. By measuring edge profiles of a fine grid (mesh 200) in the position of the k-aperture (Fig. 1) we cross-checked that the entrance and exit optics and the analyzer itself do not set the limit to k-resolution. This measurement revealed that 500 pixels were resolved along the image diagonal (setting with field-of-view dia. 1.5 Å$^{-1}$). The resulting value of 0.003 Å$^{-1}$ is an upper limit of the resolution of the entire electron optical system except the objective lens. This measurement has been done with threshold excitation by the laser (cf. section III.D). It is clear from theory (see, e.g. [1], [98] and references therein) that ultimate k-resolution requires a high extractor field or a very small photon spot (10 μm region), yielding an excellent depth of focus. Our lens system including extractor is identical to that in ref. [1], so we expect a total k-resolution of ∼ 0.005 Å$^{-1}$.

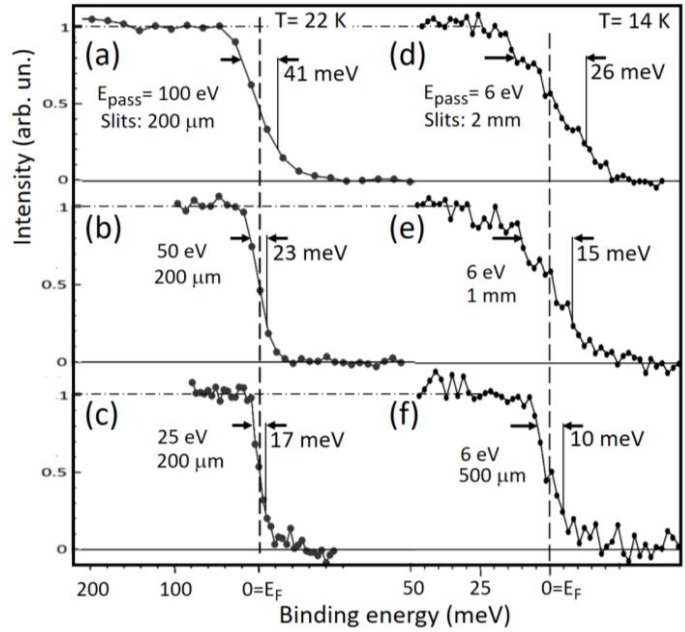

**FIG. 2.** Energy resolution for various pass energies and slit widths, measured using He I (21.22 eV) radiation. (a,b,c) Au Fermi-edge spectra measured for the smallest slit width of 200 μm at pass energies of 100, 50 and 25 eV, respectively. (d,e,f) Same for a pass energy of 6 eV and entrance apertures of 2 mm, 1 mm and 500 μm dia., respectively. Sample temperature 22 K (a-c) and 14 K (d-f); the estimated widths (error ±1 meV) contain a contribution from the thermal broadening of the Fermi edge and the bandwidth of the He I line.

### III.C. Non-isochromaticity ($\alpha^2$-term) as function of the angular filling of the analyzer

The considerations on the phase-space acceptance in Section II.A made clear that MMs offer many degrees of freedom for the optimisation of the total recording efficiency, especially when using the *dispersive-plus-ToF* operation mode. We studied the non-isochromaticity ($\alpha^2$-term) in a wide range of pass energies from $E_{pass}$= 6 to 660 eV, lateral magnifications from M = 3 to 30 and angular fillings from $\alpha_{max}$ = 0.9° to 7°. The magnification M refers to the Gaussian image in the entrance plane with respect to sample coordinates. $\alpha_{max}$ is the angle (half cone angle) which defines the non-isochromaticity, cf. eq. (1) and Fig. 1. We distinguish between the *small-angular-filling* regime, where the non-isochromaticity is practically negligible and the *large-angular-filling* regime, where this effect needs to be corrected. The border between the two regimes is not sharp, but depends on the desired resolution, i.e. on pass energy and slit widths.

Figure 3 shows results from a series of measurements recorded at eight pass energies (given in the centre column). Most measurements were taken for evaporated Au-films on Re(0001), because they are easy to produce and their *sp*- and *d*-bands show good contrast (film quality and thickness varies between the panels). Only a few measurements [(g-j) and the 25 eV results in (k,l)] have been recorded for clean Re(0001). Figs. 3(a-f) correspond to the *small-angular-filling* regime, the top row shows $k_x$-$k_y$ momentum patterns at the Fermi energy (a) and $E_B$= 20, 40 and 60 meV (b, c and d, respectively). With increasing binding energy, the images get brighter but there is no visible curvature of the isoenergetic surfaces. The first two columns show pass-energy series of sections in the non-dispersive plane $E_B$-$k_x$ (e), and dispersive plane $E_B$-$k_y$ (f). Fermi edges are marked by dashed lines and values of $\alpha_{max}$ are denoted in column (f). Please notice that all band-structure plots have the same $k_{\parallel}$-scale from -2 to +2 Å$^{-1}$ (range of emission angles -90° to +90° for He I excitation), but the angular filling $\alpha_{max}$ of the analyzer is varied between 0.9° at $E_{pass}$= 660 eV and 4.1° at $E_{pass}$= 6eV using the zoom lens. In a few panels a larger binding-energy range is shown, extending to the *d*-band onset (the *d*-band region is shown with different grey level, accounting for the higher intensity). In some panels the bands look blurred, for the high pass energies due to insufficient energy resolution and in some cases because of aged Au-films. Close inspection of the Fermi-edge regions reveals that in both columns (e,f) there is no visible curvature throughout the entire range of pass energies.

The right half of Fig. 3 shows the analogous results for the *large-angular-filling* regime. The significance of the $\alpha^2$-term is already visible in the $k_x$-$k_y$ patterns (g-j): At $E_F$ (g) the intensity exhibits a horizontal band and the upper and lower rims of the field-of-view stay dark. At $E_B$= 20 meV (h) the band widens, at 40 meV (i) the upper and lower rims are still darker than the centre and only at $E_B$= 80 meV (j) the full field-of-view is bright. This fingerprint of the $\alpha^2$-term is quantified in the $E_B$-$k_{x,y}$ sections, columns (k,l). The $k_{\parallel}$-scale is the same as in columns (e,f), but now the angular filling $\alpha_{max}$ is varied between 3° at $E_{pass}$= 660 eV and 6.5° at $E_{pass}$= 6 eV. In the non-dispersive direction (k), the Fermi edge still looks straight because the non-isochromaticity does not act on the angle $\beta$. The cuts in dispersive direction column (l), however, reveal substantial curvatures of the Fermi-energy cutoff, as predicted by eq. (1).

We quantify the non-isochromaticity as apparent energy difference $\Delta E$ between the positions of the Fermi edge for $\alpha$= 0 ($k_{\parallel}$= 0) and $\alpha_{max}$ (values given in column (l)). Measured values for $\Delta E$ are listed on the right-hand side of column (l). Values for $\alpha_{max}$ have been measured by switching off all lenses in the exit optics (Fig. 1), leading to a slightly blurred k-image on the detector which directly gives $\tan\alpha_{max}$ as ratio of image radius and distance from exit slit. The values were cross-checked by ray simulations and by proving the consistency with eq. (6). Corresponding magnifications range from M= 10 to 22 (3 to 14) for the small- (large-) angular-filling regime.

For $E_{pass}$= 50 eV, the non-isochromaticity $\Delta E$= 495 meV at the large filling angle $\alpha_{max}$= 6.0° (half cone angle) can be directly compared with the value of 100 meV at 2.6° measured using a smaller hemisphere with 150 mm path radius [9]. Taking into account the quadratic dependence on $\alpha$, these values are compatible; the difference of 8 % lies within the error limits of the angular measurements. The *relative non-isochromaticity* $\Delta E/E_{pass}$ at a given angle $\alpha$ should not depend on the size of the hemisphere, because at a given pass energy both, energy dispersion and change of radius $R_\pi - R_0$ (eq. (1)), scale with the size of the HSA. Neglecting fringe-field corrections, $\Delta E/E_{pass}$ is expected to be constant; ref. [9] gives a value of 3x10$^{-3}$ for $\alpha_{max}$= 3°. The present results agree with this value for pass energies up to 50 eV.

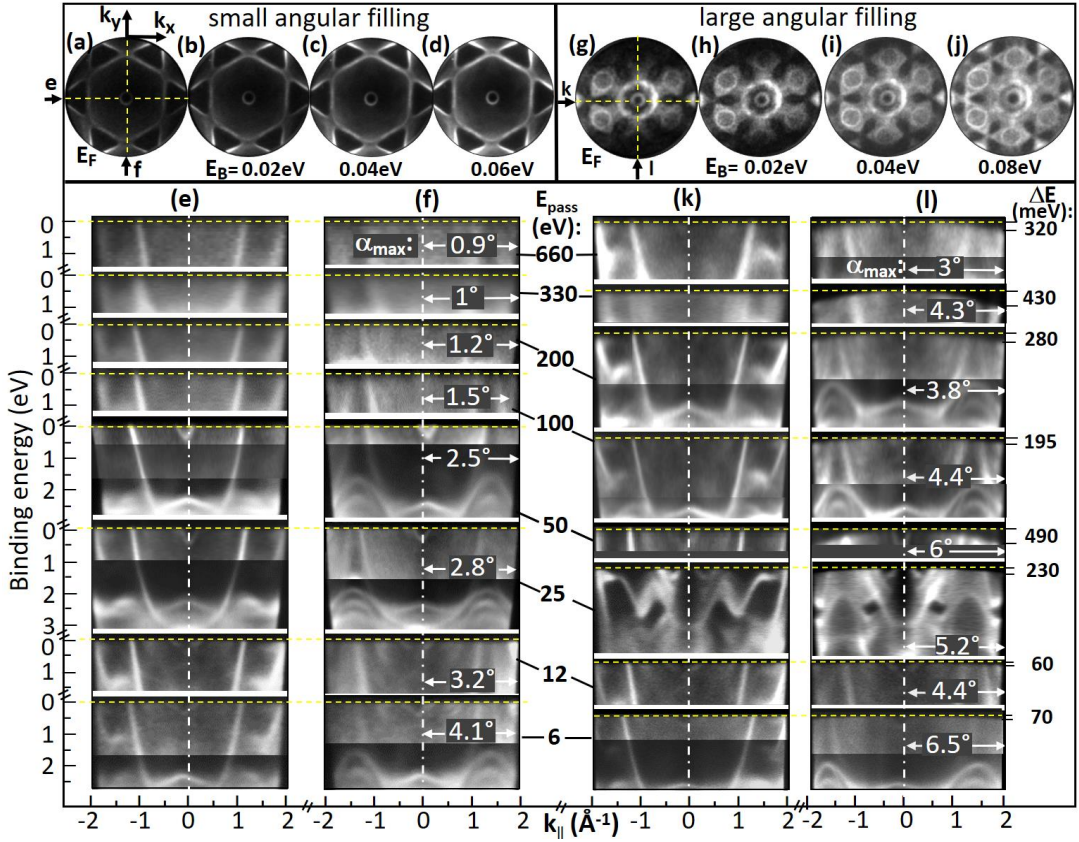

FIG. 3. Study of the non-isochromaticity ($\alpha^2$-term) for pass energies between 6 and 660 eV and small and large angular filling of the analyzer; most results were taken for evaporated Au films, a few for clean Re(0001) ((g-j) and the 25 eV result in columns (k,l)). All sections cover the same $k_\parallel$-range (radius 2 Å$^{-1}$, ± 90° on the sample surface), but recorded with different angular filling $\alpha_{max}$ in the analyzer entrance plane. The top row shows $k_x$-$k_y$ momentum patterns at different binding energies for small (a-d) and large angular filling (g-j). The four columns display pairs of $E_B$-vs-$k_\parallel$ sections in the x-z plane (angle $\beta$, columns (e,k)) and y-z plane (angle $\alpha$, columns (f,l)). Columns (e,f) show the data for small angular filling, columns (k,l) for large angular filling. Values of $\alpha_{max}$ are given in the panels; pass energies are stated in the centre. The non-isochromaticity values, quantified as apparent energy difference $\Delta E$ between the positions of the Fermi edge for $\alpha$= 0 ($k_\parallel$= 0) and $\alpha_{max}$ are given at the right side of column (l).

At higher pass energies our measured ratios $\Delta E/E_{pass}$ are significantly smaller. We tentatively attribute this deviation to the action of the Jost plates, which are individually optimised at each pass energy. It seems that shaping the fringe fields can counteract non-isochromaticity. The deviations from a constant ratio could be connected with a slight angular tilt of the incoming beam when passing the entrance slit. We could not perform a systematic study because of missing technical means to tilt the incoming beam in a well-defined way. The observed deviation resembles the result of Zouros et al. [90-94] concerning the paracentric entrance condition. The importance of the fringe-field correction on the total analyzer performance is also discussed by Wannberg [78]. Presently it is not clear, however, whether this reduction of the $\alpha^2$-curvature at $E_F$ comes at the expense of k-resolution.

In practical work, settings with small $\alpha_{max}$ (demanding high lateral magnification) and hence negligible $\Delta E$ can be used if the photon spot is small. However, the non-isochromaticity can be corrected numerically using the empirical curvature of an isoenergetic surface like the Fermi edge, as first demonstrated in [9]. The present study reveals that filling angles up to $\alpha_{max} = \pm 7°$ are possible in order to optimize the intensity in case of a very large photon footprint. In work at a Synchrotron beamline, the reference curve for correction of the $\alpha^2$-term can be measured very rapidly using a (dispersion-free) narrow core-level signal. At fixed settings of the analyzer, this reference curve stays constant.

### III.D. Temporal spread and its correction in k-resolving detection

Conventional hemispherical analyzers for high-resolution ARPES employ position-sensitive electron detectors in the exit plane. Kinetic-energy and emission-angle intervals are recorded simultaneously in a 2D-array $I(E_{kin},\theta)$. The $E_{kin}$-interval comprises typically 10% of the pass energy and the angular range for θ (polar emission angle on the sample) is ± 7.5° for high angular resolution and two times larger for the wide-angle mode. The emission angle θ at the sample surface should not be mixed up with the filling angle $\alpha_{max}$ of the analyzer. As mentioned above, the conventional HSAs have reached excellent performance close to the theoretical limits. The detailed work in [77,79] reveals, however, that their time resolution is strongly limited (time spread up to tens of ns) due to the different transit times on the Kepler orbits.

A way out of this dilemma is to place the detector not in the exit plane, where for each energy a real-space image of the entrance slit exists (unifying rays from all angles α), but in a Fourier plane. In the Fourier image the arriving electrons are sorted by their entrance angles α and β (see plane FP5 in Fig. 1). In the analyzer entrance and exit apertures the momentum pattern is encoded in terms of these angles, hence each Kepler orbit corresponds to a certain arrival position in the final k-image on the detector. In turn, the arrival time for the electrons on each individual orbit can be directly observed (and corrected) in experiments with pulsed excitation sources and time-resolving detector. According to theory, the transit times should be strictly deterministic, i.e. at each energy they are given by the angular coordinates of the rays in the entrance slit.

We studied the transit-time spread for pass energies between 50 and 660 eV. A picosecond-laser (400 nm, 50 ps pulse width) served as test source for pulsed excitation, a DLD [23,96] for time-resolved recording. The result obtained for a cesiated Au film is shown in Fig. 4. Due to the small photon energy (3.1 eV) the momentum pattern of the cesiated film is only 1 eV wide and has a rather small diameter of only 1 Å$^{-1}$. The $k_x$-$k_y$ pattern, Fig. 4(a), shows a weak residue of the Shockley state, strongly quenched by Cs adsorption. The full photoemission paraboloid for laser excitation is shown in the $E_B$-vs-$k_x$ section (b), measured in the conventional way by stepwise scanning of the energy (see first para. of II.B). This data paraboloid $I(E_B,k_x,k_y)$ is defined by the photoemission horizon (emission angle θ= 90° at the sample surface) and the Fermi edge. For time-resolved detection the coarse binding energy $E_B$ is defined by the analyzer setting and the DLD records the transit time τ as fourth "coordinate", yielding data arrays $I(E_B,\tau,k_x,k_y)$. The width of the transmitted energy band depends on pass energy and entrance / exit apertures. The transit-time spread Δt of the rays in the analyzer is directly visible in τ-vs-$k_{\|}$ sections through this data array.

The τ-vs-$k_x$ sections at pass energies of 200 and 660 eV (Fig. 4(c,d)) confirm that the time-spread along the non-dispersive direction is zero within the experimental uncertainty, as expected for ideal spherical symmetry (angle β in top right inset of Fig. 1). For finite entrance and exit apertures it is likely that higher-order terms act on the time spread in non-dispersive direction. However, we are not aware of theoretical work on a time spread as function of the angle β.

The sequence of sections along the dispersive direction τ-vs-$k_y$ at pass energies between 50 and 660 eV (Fig. 4(e-l)), reveals sizeable values of Δt between ∼10 ns for $E_{pass}$= 50 eV (f) and ∼2 ns for 660 eV (l). These measured results are in good agreement with Δt = 7.4 and 2.0 ns for α= 1°, predicted by eq. (5). Moreover, Figs. 4(e-l) confirm that Δt is a linear function of $k_y$ and hence a linear function of α. As anticipated, the third-order term in eq. (5) is negligible at the given conditions. The patterns Fig. 4(f,h,j) have been purposely widened via the zoom lenses in order to cover a larger range of $k_{\|}$. The slight curvature is most likely an artefact of this electron-optical shaping.

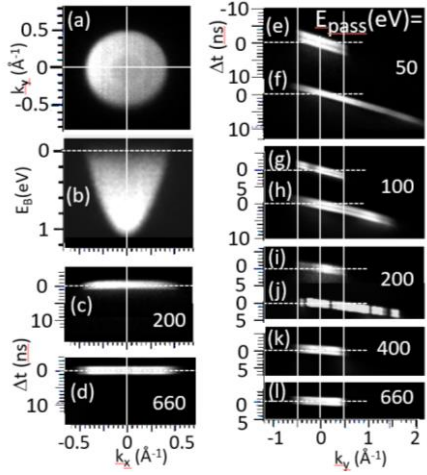

**FIG. 4.** Transit-time spread Δt measured as function of pass energy between 50 and 660 eV. (a) and (b), $k_x$-$k_y$ and $E_B$-vs-$k_x$ momentum patterns, respectively, recorded for $E_{pass}$= 200 eV and 500 μm aperture by scanning the analyzer energy with the ToF switched off. (c,d) Time spread Δt as function of parallel momentum $k_x$ (along the non-dispersive direction) for $E_{pass}$= 200 eV and 660 eV. (e-l) Time spread as function of parallel momentum $k_y$ (along the dispersive direction) for five different pass energies. The short patterns (c-e,g,i,k,l) show the temporal structure of one energy slice from the distribution in (a,b), whereas the long patterns (f,h,j) have been stretched via the electron optics in order to increase the filling angle, simulating a larger k-field-of-view.

The central result of Fig. 4 is that the time lag of electrons travelling on different Kepler orbits depends in a well-defined way on the momentum component $k_y$. This opens a route towards a numerical correction. We propose the procedure for transit-time correction as illustrated in Figure 5. Fig. 5(a) shows an $E_B$-vs-$k_y$ section through the as-measured momentum pattern, recorded at a pass energy of 200 eV with an entrance aperture of 500 μm. This yields a transmitted bandwidth of 220 meV, showing up as tilted disc in the (τ,$k_x$,$k_y$) parameter space. In the τ-vs-$k_y$ section Fig. 5(a) the Fermi edge ($E_B$= 0) appears as sharp upper border of the tilted bright stripe. Due to the temporal dispersion of the ToF section (drift energy 20 eV) the Fermi edge appears much sharper than the 220 meV bandwidth, obtained for $E_{pass}$= 200 eV and 500 μm aperture. However, the ToF spectrum "rides" on the strongly tilted baseline due to the transit-time behaviour.

The schematic 3D-sketch in Fig. 5(b) explains, how the parameters binding energy $E_B$ and transit time τ are "intermixed" in terms of a tilted energy scale. The *transit time artefact* in the ToF patterns can be eliminated by data treatment, with the slope of the tilted plane serving as fit parameter. After the correction of the time scale, linear in the coordinate $k_y$, the processed experimental pattern (Fig. 5(c)) looks like typical patterns recorded by a conventional ToF-MM [103]. Sketch (d) shows that now the two effects, transit time lag and ToF dispersion are completely disentangled, so that the scale of τ is a measure of kinetic energy. The paraboloid is truncated at the bottom due to the pre-selected energy band passing the analyzer exit aperture.

Fig. 5(e) shows the result of a conventional measurement taken with the same analyzer setting but without ToF recording. Comparison of this standard measurement with the measurement in the ToF hybrid mode (c) demonstrates the substantial enhancement of resolution. Despite the small photon energy, the measurement with the ps-laser gives a clear proof-of-principle of the gain of energy resolution for a given setting of the hemispherical analyzer. In band mapping experiments at Synchrotron beamlines or other pulsed photon sources the energy is scanned stepwise through the region of interest (via variation of the sample bias, keeping all other potentials fixed). Each counting event registered by the DLD is saved in a histogram memory yielding the 4D data array $I$ ($E_B$,τ,$k_x$,$k_y$). First, the tilt of the entire distribution due to the different transit times is eliminated using the empirical linear function Δt($k_y$), yielding the corrected array $I$ ($E_B$,$τ_{corr}$,$k_x$,$k_y$). Second, the value of $τ_{corr}$ for each individual event is used for a precise determination of the corresponding binding energy, according to $E_B^* = E_B + ΔE(τ_{corr})$. Here, $E_B$ is the binding energy set by the analyzer electronics at the instant of the counting event and ΔE is the energy correction resulting from the time-of-flight $τ_{corr}$ of the electron through the drift section of the ToF analyzer. In the final data array I ($E_B^*$,$k_x$,$k_y$) the energy resolution is substantially (1 to 2 orders of magnitude) higher than in a measurement without the ToF-booster. For each (analyzer-) setting of $E_B$ the measurement of $τ_{corr}$ yields the precise time slice within the energy bandpass. In addition to the transit-time spread due to the different path lengths the electrons within the transmitted band pass experience a slight temporal dispersion when travelling inside the HSA (path length ~ 0.7 m). In the hybrid mode (typical pass energies several hundred eV) this dispersion is small in comparison with the dispersion of the ToF section (drift energies 10 - 30 eV, length 0.5 m) and can be corrected in the data processing.

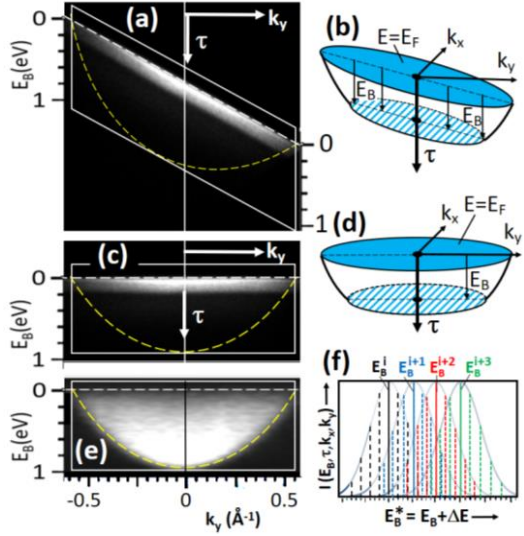

**FIG. 5.** Elimination of the transit-time spread for $E_{pass}$= 200 eV, entrance aperture 500 μm and drift energy in the ToF section 20 eV; sample temperature 22 K. (a) As measured distribution τ-vs-$k_y$ and (b) schematic sketch of data array $I$ ($E_B,τ,k_x,k_y$), showing the intermixing of time coordinate τ and binding energy. (c) Same distribution like (a) but after numerical correction using an empirical function linear in $k_y$. (d) Schematic sketch after correction, now $E_B$-scale is directly related to time-of-flight τ. (e) Same data array measured for $E_{pass}$= 200 eV and 500 μm aperture without ToF analysis by scanning the sample bias. Note the strongly reduced resolution, compare Fermi edges in (c) and (e). (f) Data recording scheme of the *dispersive-plus-ToF* hybrid mode: The HSA scans $E_B$ stepwise (i, i+1, i+2,...) and for each step the DLD records the events with a higher energy resolution; time slices indicated as coloured dashed lines. In (a,c) sample bias was fixed, whereas in (e) it was scanned.

This data recording scheme is illustrated in Fig. 5(f). The Gaussian curves show the transmitted energy bands during stepwise scan of the binding energy $E_B$ (i, i+1, i+2...). The DLD software sorts each counting event into a histogram memory with the binding energy corrected by the term ΔE resulting from the ToF measurement. The intervals of ΔE are indicated by dashed coloured lines. For this mode the small-angular-filling regime is favourable because both, transit time-spread and $α^2$-aberration, are smaller. The quadratic term ($α^2$-curvature) is superimposed to the linear term (transit-time spread ΔE), so both could be corrected independently. The precision of this correction is given by the pulse width of the excitation source convoluted with the time resolution of the detector.

Concluding this section, we consider the gain in recording efficiency when using the *dispersive-plus-ToF* hybrid mode. It results from the fact that $E_{pass}$ can be strongly increased when using this mode. If N is the number of resolved time slices, to achieve the same energy resolution, $E_{pass}$ can be increased by a factor of N without changing the analyzer slits when switching from the mode without ToF analysis to the hybrid mode. Observed k-field-of-view ($k_{\parallel,max}$) and angular filling of the analyzer ($α_{max}$) are kept fixed. Note that at $N*E_{pass}$ the magnification M is reduced by a factor of $N^{1/2}$ due to Eq. (6). As long as the total photoelectron signal entering the analyzer is proportional to $M^{-2}$ (cf. discussion in III.A) the intensity entering the entrance aperture increases by a factor of N and the total energy bandwidth also increases by a factor of N. As a result, the intensity per time slice is N times higher at $N*E_{pass}$ and the ToF device records N time slices in parallel, yielding a total gain of $N^2$. When the image of the photon spot lies completely within the entrance aperture, the change of magnification does not increase intensity and the total gain will be just linear in N. In the high-resolution regime this change from quadratic to linear dependence on N occurs at a photon spot size of ~10 μm.

We can quantify the expected gain at a Synchrotron source running at 500 MHz photon-pulse rate. The DLD (~70 ps time resolution [96]) will enable us to resolve 20-25 time slices in the 2 ns interval between adjacent pulses. At a pass energy of 400 eV and apertures of 500 μm the bandpass is 440 meV. The DLD cuts this band into time-slices of 20-25 meV width, fitting perfectly to the photon bandwidth of high-performance soft X-ray beamlines [100,101]. Without ToF analysis, this resolution would require a pass energy of 20 eV, connected with a loss factor of $20^2$ in recording efficiency. A key application of the ToF hybrid mode will be spin-resolved experiments, because the imaging ToF spin filter also benefits from this gain factor [65].

## IV. RESULTS AND DATA PROCESSING

As example for full-half-space k-mapping we have chosen two cases of quantum-well (QW) states in multilayers with emphasis on the numerical processing of very weak signals coexisting with strong band features. An excellent introduction and review of early work on QW states in Au (and Ag) multilayers is given by Chiang [104]. Matsuda et al. [105] observed two distinctive types of photon-energy dependence of the QW state binding energies, with oscillatory shifts for $h\nu < 15$ eV being a fingerprint of the quantized *sp*-type final-state band in contrast to the continuum-like *d*-band for $h\nu > 20$ eV. QW states of metal-on-semiconductor junctions [106] revealed aspects beyond the particle-in-a-box model [107] and showed the relevance of correlations with substrate electrons, visible in the effective mass $m_{eff}$, ranging from 0.2 to 0.3 $m_0$ ($m_0$ free electron mass) [108,109]. The relative photoemission intensities of the QW states can exhibit dramatic variations with film thickness and photon energy [104]. This is attributed to oscillations of the dipole matrix elements that govern the transition from the QW state to the photoemission final state, depending on the relative phases.

The rapid data acquisition of a MM is ideally suited to track the dispersion and intensity of QW states and their thickness dependence. Figure 6 shows a collection of results for Au multilayers on Re(0001). The top row illustrates some ways of numerical data treatment, offered by full-field imaging. The unfiltered He-spectrum of the lamp contains an admixture of He-Iβ (23.08 eV), about 3% [99] of the main line He-Iα (21.22 eV). This contribution can be eliminated for the k-pattern at $E_F$, by subtracting the He-Iβ pattern Fig. 6(b), measured 50 meV above the edge, from the as-measured He-Iα pattern (a). The difference $k_x$-$k_y$ image (c) shows the pure He-Iα momentum map of the Au *sp*-band, the Shockley state (S) in the centre and a series of quantum-well (QW) states. The "He-Iβ artefacts" (arrows in (a,b)) are eliminated in (c). Notice that pattern (b) corresponds to $E_B$= 1.86 - 0.05 = 1.81 eV due to the higher photon energy and 50 meV kinetic-energy difference in the two measurements. Similarly, the change of the QW states with varying thickness can be emphasized by subtracting two patterns taken for different thicknesses. Such an example is shown in Fig. 6(d), the difference image of the energy sections at $E_F$ for the 6 and 10 monolayer Au-films.

Figures 6(e,f) show the $E_B$-vs-$k_{x,y}$ sections along the $\bar{\Gamma}$-$\bar{M}$ (e) and $\bar{\Gamma}$-$\bar{K}$ direction (f), see BZ in (c). In order to emphasize the weak QW-states coexisting with the much stronger *sp*-band and Shockley state (S), patterns (e,f) are numerically treated: First the intensity is normalized by dividing each energy slice by its integrated intensity value, thus homogenizing the intensities along the "$E_B$-coordinate". Second, the resulting k-distribution is displayed using a logarithmic grey scale, thus emphasizing the weak features in one energy slice. The two columns to the right show $k_x$-$k_y$ sections at different binding energies (as marked by arrows) for film thicknesses of 10 (g-j) and 6 monolayers (k-n). The ring-shaped QW-states marked by arrows in (g,k) are different in the two columns because they depend on film thickness. These results agree perfectly well with literature data [104-109]. In addition, further QW states are visible close to the $\bar{M}$-point in the gap between the *sp*-bands of the first and second BZ (red arrows in (e)). Figs. 6(o-q) show $k_x$-$k_y$ sections in the region of the *d*-band onset.

The large dynamic range accessible via two-step data processing (homogenizing along $E_B$ and log-scale display) is demonstrated in Fig. 6(r). The faint band features close to the $\bar{M}$-point marked by ellipses are one and two orders of magnitude weaker than the *sp*-band and *d*-band, respectively. This cut runs through the top of the *d*-band at $\bar{M}$. Next to the well-known series of paraboloid-shaped QW-states (e-n), we found a second appearance of the electronic confinement in the Au multilayers, shown in Fig. 6(s,t). A series of quantizes states without dispersion appear in terms of horizontal stripes in the region close to the photoemission horizon (i.e. at almost grazing emission).

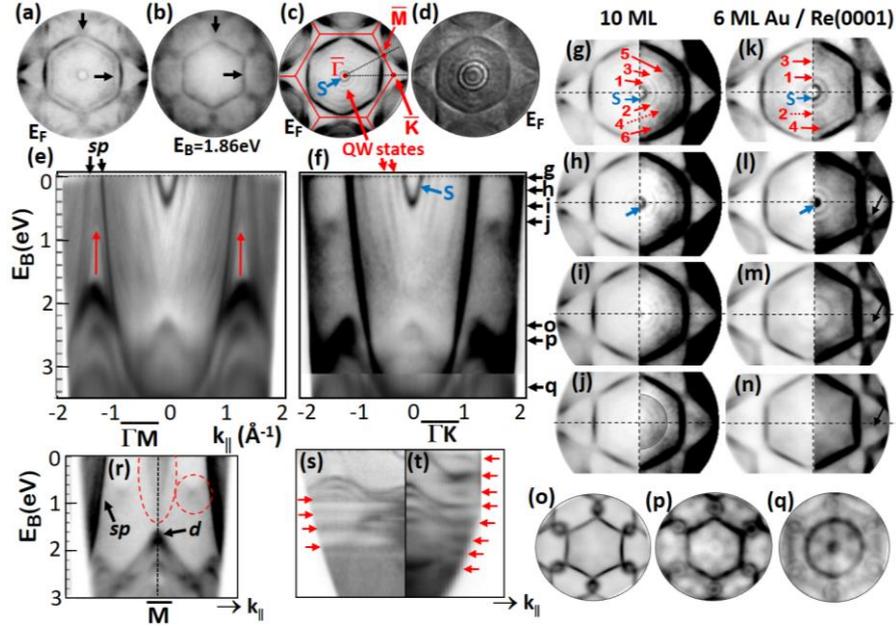

**FIG. 6.** Quantum-well states in ultrathin Au(111) films on Re (0001), recorded at 22 K. (a-d) Examples for data treatment via pixel-by-pixel processing of two images. Removal of a contribution of the He-Iβ line (23.08 eV) from a raw image recorded at the Fermi edge (a) by subtracting a pre-edge image taken 50 meV above $E_F$, i.e. at $E_B$= 1.81 eV (b), yielding the clean He-Iα (21.22 eV) $k_x$-$k_y$ pattern (c). The surface Brillouin zone with high-symmetry points $\bar{\Gamma}$, $\bar{M}$ and $\bar{K}$ is marked in (c). Ratio image (d) of two QW-state patterns for different coverages emphasizes the differences. (e,f) $E_B$-vs-$k_\parallel$ sections along $\bar{\Gamma}$-$\bar{M}$ and $\bar{\Gamma}$-$\bar{K}$, respectively, after intensity renormalization on a logarithmic grey scale. (g-j) and (k-n) series of $k_x$-$k_y$ patterns for 10 and 6 monolayers (ML) of Au taken at the binding energies marked by arrows in (f); QW-states marked by arrows in (g,k). (o-q) show $k_x$-$k_y$ patterns taken in the onset region of the *d*-band. (r) Section with high dynamic range of 2 orders of magnitude displayed on a logarithmic grey scale. (s,t) $E_B$-vs-$k_\parallel$ sections showing series of quantized dispersion-free states (arrows) for 5 and 12 ML, respectively.

Xe(111) films of well-defined thickness and homogeneity are easy to produce by dosing at sufficiently low substrate temperature, leading to Van-der-Waals *fcc* crystals with (111) surface orientation. The relativistic band structure was calculated as early as 1970 by Rössler [110] using the Green's function method. The 5*p*-derived bands and their double-group symmetries were confirmed by spin-polarized photoemission [111]. The 5*p*-level of a free Xe atom shows a spin-orbit splitting of 1.3 eV. In the cubic crystal field the $5p_{3/2}$ state splits into its $|m_j|$= $3/2$ and $1/2$ sub-levels (double-group notation along the Γ-L direction $\Lambda^3_{4,5}$ and $\Lambda^{1,3}_6$, respectively), sometimes termed heavy- and light-hole bands due to the different effective masses. The $5p_{1/2}$ state (notation $\Lambda^{3,1}_6$) is spherically symmetric and does not split in a paramagnetic material. In the following we label these bands 1, 2 and 3 with increasing binding energy. Groups in Bonn and Kassel discovered QW states in Xe multilayers [112-114], which show up as a coverage-dependent multitude of peaks in the binding-energy range between ~5 eV and 8.3 eV.

Figure 7 shows a collection of results for Xe on Re(0001) for a *film of 3-4 layers* (a-n), for a *single monolayer* (o-q) and with *increasing coverage* (r-z). The last pattern (w-z) was measured on a Mo(110) substrate. Xe was dosed at 45 K and the coverage was estimated assuming a sticking coefficient of 1. The monolayer was annealed at 58 K, desorbing the multilayer. The lattice constant of solid Xe (6.13 Å) is 50% larger than that of Au (4.07 Å). In turn, the k-field-of-view with He I excitation extends almost to the centres of the next Brillouin zones (see Fig. 7(a)). All data in Figs. 7(a-n) are from a single scan, recorded within a few hours (nominal resolution 27 meV, energy steps 20 meV).

Figs. 7(a-j) show selected $k_x$-$k_y$ sections in the region of the Xe 5*p* band (a-g) and below the 5*p* band complex (h-j); corresponding energy positions are marked by arrows at the left rim of (k). All sections show pronounced band features with clear sixfold symmetry (except for dichroism effects in the intensities), validating the high quality of the multilayer films. The $E_B$-vs-$k_\parallel$ sections corresponding to the $\bar{\Gamma}$-$\bar{M}$ and $\bar{\Gamma}$-$\bar{K}$ directions (Figs. 7(k) and (l), respectively) reveal a number of Xe-derived dispersing bands throughout the entire region $E_B$= 5-14.5 eV. Curves labelled 1-3 in Fig. 7(k) show the calculated bands from [110], adapted to the experimental binding energy. Topmost band 1 appears at a binding energy of 5.1 eV at the $\bar{\Gamma}$-point (Fig. 7(a)) and only 200 meV below two further band heads appear at $\bar{\Gamma}$ and at the six $\bar{M}$-points (b). Band 2 shows a pronounced sixfold feature in section (c), taken slightly above the bottom of the $p_{3/2}$-derived band complex. The top of the $p_{1/2}$-band (3) appears at $E_B$= 6.8 eV and is obviously split into two band heads (d), looking similar to cut (b). Band 3 disperses downwards towards its minimum at $\bar{M}$, in good agreement with the calculation. In the vicinity of its bottom very pronounced sixfold-symmetric k-patterns appear (Figs. 7(e-g)). The region of bands 1 and 2 (5-6.6 eV) clearly shows more than two bands, and also band 3 is split into three sub-bands (arrows in (k,l)). At the lower rim of the 5$p$-complex ($E_B$= 7.95 eV) we observe a narrow band with energy width of ~50 meV at $\bar{M}$, marked as band 3 in the section with linear intensity scale (m). In the intensity spectrum (n) this band shows up as sharp peak. In the small interval between the intensity maximum of band 3 (7.7 eV) and the region below the band bottom (8 eV) the intensity drops by 2 orders of magnitude.

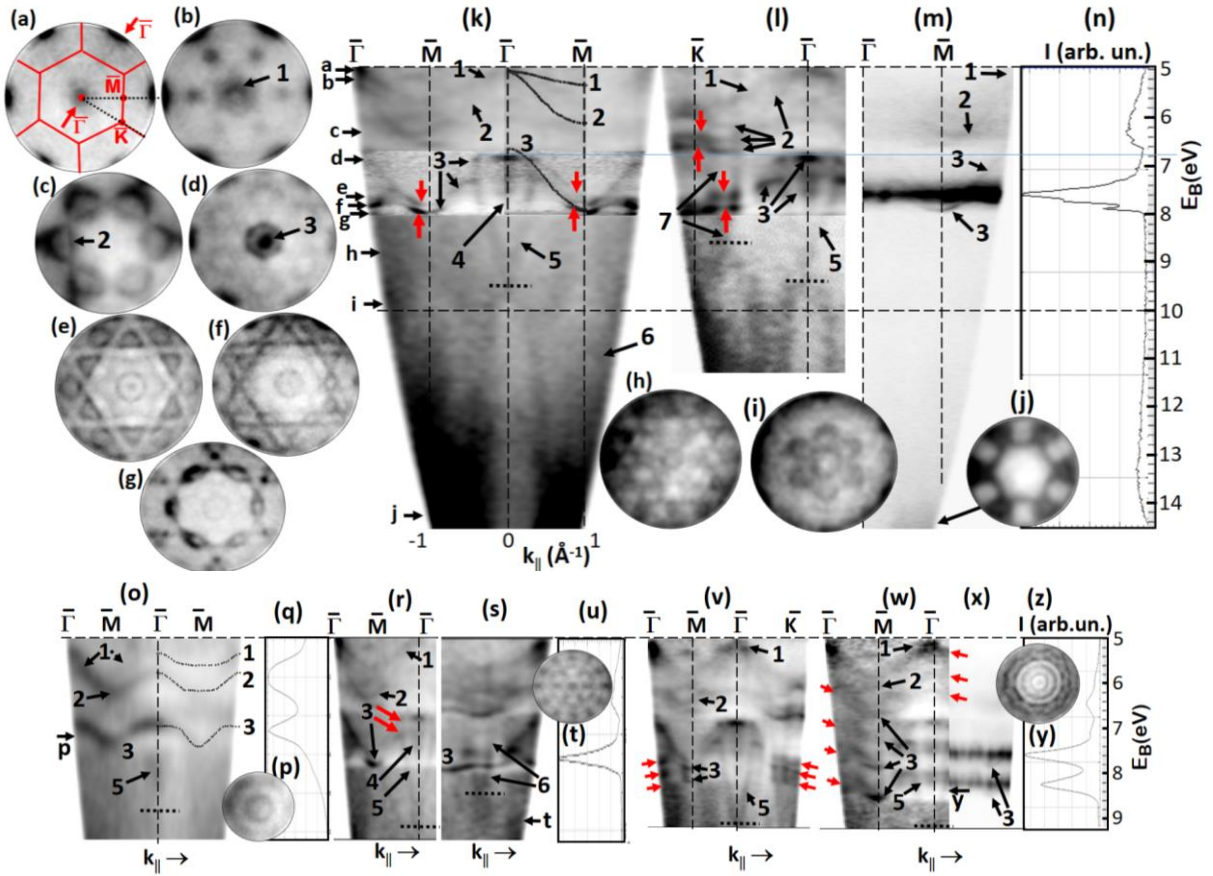

FIG. 7. (a-n) Xe 5$p$ valence bands in a Xe film of 3-4 ML on Re(0001). (a-g) $k_x$-$k_y$ constant-energy patterns in the region of the 5$p$-bands and (h-j) below the 5$p$-band complex. The surface Brillouin zone with high-symmetry points $\bar{\Gamma}$, $\bar{M}$ and $\bar{K}$ is marked in (a). (k,l) $E_B$-vs-$k_\parallel$ sections along the $\bar{\Gamma}$-$\bar{M}$ (k) and $\bar{\Gamma}$-$\bar{K}$ directions (l). Full curves labelled 1-3 in (k) show the calculation of Xe bulk bands (along $\Lambda$) by Rössler [110]. Bands 4-7 appear below the bottom of the 5$p$-band. (m) same as (k) but on a linear intensity scale; (n) corresponding intensity spectrum. (o-z) Thickness dependence of bands in Xe-films with thickness 1 ML (o-q), 2-3 ML (r-u) and ~6 ML on Re(0001) (v) and ~9 ML on Mo(110) (w-z). Dotted curves labelled 1-3 in (o) show a tight-binding calculation of Xe-monolayer bands by Kambe [116]. On Re(0001) there is only one lateral orientation of the Xe film, whereas on Mo(110) two orientations exist (rotated by 30°), leading to an apparent 12-fold symmetry (y). $E_B$-vs-$k_\parallel$ sections shown on a logarithmic grey scale after intensity renormalization, except for (m,x). Red arrows mark bands which are split or replicated.

Surprisingly, even below the 5*p*-band complex down to $E_B$= 14.5 eV we still find several Xe-derived bands (4-7) with sixfold-symmetric k-patterns (h-j) and some crossing points (dotted horizontal lines in (k,l,r,s,v,w)). These bands cross the 5*p*-complex, causing strong intensity modulations in the 5*p* bands (e.g., band 7 in Fig. 7(l)).

In order to emphasize weak signals coexisting with strong ones the data arrays were treated by two-step processing (normalization of all slices to equal average intensity plus logarithmic scale), as described for Au films (Fig. 6). The unprocessed data stacks and spectra (Fig. 7(m,n,x,z)) reveal that the patterns are largely dominated by one prominent peak at $E_B$= 7.7 eV (m,n) and two such peaks at $E_B$= 7.6 and 8.3 eV (x,z). The bottom part (Fig. 7(o-z)) shows patterns for 4 different coverages, starting with a single saturated hexagonal monolayer (o-q). The Xe 5*p* dispersion in a monolayer was first measured by Horn et al. [115], who found evidence that the splitting is due to lateral Xe-Xe interactions. The tight-binding calculation of Kambe [116] (dotted curves in Fig. 7(o)) agrees well with our measurement. Band 1 is well visible in the 2$^{nd}$ BZ but disappears at the central $\bar{\Gamma}$-point (centre of (o)), in agreement with a previous observation (see Fig. 13 in [117]). Figs. 7(r-u), (v) and (w-z) show results for increasing film thicknesses of 2-3, 6 and 9 ML, respectively. It is eye-catching that the patterns change substantially. We do not observe series of QW states with parabolic dispersion, but rather the bands appear "repeated" as indicated by the red arrows. In the 9 ML film (w) a series of repetitions of band 3 with a spacing of ~700 meV is visible, being an appearance of QW states. These QW states are weaker for bands 1 and 2 (top right of (w)), which might reflect the fact that the $p_{1/2}$-state (band 3) is spherically symmetric but $p_{3/2}$ is not. Fig. 7(w-z) was recorded for Xe on Mo(110), in order to exclude that band 5 is an effect of the Re(0001) substrate.

Presently there are no calculations that can explain these findings. The small band splittings in the *few-layer regime* can be tentatively explained by layer-dependent image-charge screening, a special appearance of quantum size behavior. In an insulator like Xe the screening of the photo-hole by the conduction electrons of a metal substrate decreases stepwise in the sequence 1$^{st}$, 2$^{nd}$, 3$^{rd}$, etc. atomic layers. For Xe on Pt(111) the corresponding binding energy increases by 0.54 eV between Xe$^+$ ions in the 1$^{st}$ and 2$^{nd}$ and by ~0.2 eV between 2$^{nd}$ and 3$^{rd}$ layers (cf. Figs. 18(k,m,n) in [117]). We propose that the band splittings in Fig. 7(k,l,r) originate from this quantization of the hole screening. Please notice that the first monolayer carries a substantial surface dipole, lowering the work function by ~0.5 eV. To the best of our knowledge, the repetition of entire band groups, most strikingly visible when comparing Figs. 7(m) and (x) has not been observed for Xe films before. Also the lower-lying bands 4-7 with their pronounced k-patterns (h-j,t) have not been reported previously, most likely because they are orders of magnitude weaker than the main 5*p*-bands. Surprisingly, band 5 is already indicated in the Xe ML, Fig. 7(o), i.e. in a 2D electronic system. One-step photoemission calculations are in progress which should uncover the possible presence of final-state effects. Resonant excitation into unoccupied bands might be the reason for the strong intensity modulations in the crossing regions, as discussed in [104] for Au QW states.

The measurements of Fig. 7 are a good example for the performance of full-half-space k-imaging over a large range of ($E_B$,**k**) phase space. It is not necessary to know in advance, where the interesting physics happens. Including the structure-less region between $E_F$ and $E_B$= 5 eV (not shown here), the survey measurement (a-n) comprised an interval of 15 eV, screened in steps of 20 meV. The data stack thus consists of 750 $k_x$-$k_y$ energy slices in a k-interval of the central BZ plus about half of the sixfold symmetric first repeated BZs.

## V. SUMMARY AND CONCLUSION

This paper presents a new approach for angle-resolved photoelectron spectroscopy (ARPES), combining a single hemispherical analyzer (HSA) - operated as momentum microscope - with time-of-flight parallel energy recording. HSAs constitute the vast majority of analyzers for high-performance ARPES. The main-stream mode of operation employs a position-sensitive detector in the exit plane for parallel acquisition of 2D data arrays $I(E_{kin},\theta)$. The kinetic-energy interval recorded simultaneously is typically ~10% of the pass energy and the angular range $\theta$ covers $\pm 7.5°$ to $\pm 15°$, depending on angular resolution. These analyzers are characterized by excellent energy resolution down to the sub-meV range [118,119] alongside with a relatively small interval of emission angles $\theta$ at the sample surface. The range of angles $\alpha$ in the entrance aperture of the analyzer (definition see Fig. 1) has to be kept small ($\alpha_{max}$ typically $\pm 2°$), otherwise the $\alpha^2$-term deteriorates resolution [71]. The large transit-time spread of electrons travelling on different Kepler orbits [73,75,79] is prohibitive for sub-ns time-resolved recording.

Pioneering work by Kirschner and coworkers in Halle [6,1] established a new way of ARPES using a HSA in a different mode, termed photoelectron momentum microscopy (MM). Adopting concepts of electron microscopy, the analyzer is used as imaging energy filter. The detector is placed in a Fourier plane, recording the reciprocal image (momentum image). A cathode lens with strong electric field collects a large interval of emission angles ($2\pi$ azimuth, $\theta$= 0-90° up to $E_{kin}$= 70 eV). This lens converts the photoelectron angular distribution into a momentum pattern, the HSA transmits the angular-encoded full-field 2D image $I(k_x,k_y)$, and different energies are recorded sequentially. A tandem arrangement of two HSAs with inverting lens in between was considered necessary for aberration compensation [6-8]. The k-resolution (0.0049 Å$^{-1}$ [1]) is close to best values obtained using conventional HSAs (e.g. 0.003 Å$^{-1}$ [81]) in a much smaller k-interval. Reported energy resolutions are still about an order of magnitude lower than for conventional HSAs.

Here we present the first results of MM with a large single HSA, demonstrating an energy resolution of 7.7±1 meV and k-resolution of 0.008 Å$^{-1}$ using excitation by a simple, unfocussed He I discharge source. The capability of gaining high-quality k-images is proven by mapping the well-known quantum-well states in Au multilayers and a new type of quantum-well states in Xe multilayers on Re(0001) and Mo(110). The former belief that a double-HSA is mandatory for aberration compensation was recently revisited by Tusche et al. [9], who showed by analytical treatment that refraction of the electron rays in the entrance aperture essentially eliminates the aberrations.

The key achievement of the present instrument is the implementation of an additional time-of-flight (ToF) detector behind the exit of the HSA, adopting concepts developed for the time-of-flight (ToF) MM [25]. This combination facilitates simultaneous 3D-recording $I(k_x,k_y,E_{kin})$ in energy intervals transmitted by the HSA. At first sight, ToF-recording (with a time resolution < 100 ps) seems to be at variance with the large transit-time spread (up to tens of ns, depending on pass energy and angular range). However, it is clear from theory and confirmed in our experiments that the transit time $\tau$ is a linear function of the momentum coordinate $k_y$ along the dispersive plane. This leads to a tilt of the measured $I(k_x,k_y,E_B,\tau)$ data arrays. Here $E_B$ is the binding energy set by the HSA (which is scanned stepwise) and $\tau$ is the arrival-time coordinate, measured for each individual counting event. This tilt can easily be corrected numerically, i.e. the time spread can be completely eliminated (Fig. 5). Likewise, the $\alpha^2$-curvature of energy isosurfaces can be corrected as previously shown in [9]. k-resolved detection in the Fourier plane thus circumvents the preconditions of comprehensive theoretical work on transit-time spread and $\alpha^2$-aberration. Parallel to the present instrument a new series of commercial spectrometers has been developed, which combine a similar microscopy lens system with a single HSA [120]; for first results, see [121,122].

The advantage of combining a dispersive analyzer with ToF recording is evident for photon sources with high pulse rates, e.g. storage rings running at 500 MHz pulse rate. The energy band pre-selected by the HSA can be chosen such that the number of resolved time slices within this band fits to the desired resolution. A delay-line detector with ~70 ps time resolution yields 20-25 energy slices in the 2 ns gap between the photon pulses, resolved via ToF-analysis. Tuning the drift energy in the ToF section, the width of the slices can be adapted to the photon bandwidth of the beamline. For the soft-X-ray beamline I09 at Diamond Light Source, UK (where this instrument will be installed) the photon bandwidth will be ~30 meV at 500 eV (after an upgrade of the beamline optics). ToF slicing reaches this resolution for an energy band of ~0.7 eV width, transmitted by the HSA. For 500 µm apertures this bandwidth corresponds to a pass energy of 660 eV. Without ToF analysis a resolution of 30 meV requires a pass energy of 28 eV. In the regime where the image of the photon spot in the analyzer entrance plane is larger than the entrance aperture (transmission proportional to $E_{pass}^2$), we can expect a gain in recording efficiency of >500. Future work will show how close one can get to this value in practice, because the size of the photon spot on the sample surface plays a crucial role (see detailed discussion in Section III.A). A purely ToF-based MM without HSA cannot operate at 500 MHz photon-pulse rate, because the pulse spacing is far too small to allow some hundred energy slices to be resolved. Scanning the sample bias does not work without a HSA or other dispersive element. Present bunch lengths of storage rings are ~50 ps. The new low-emittance machines are designed for somewhat larger bunch lengths, which will reduce the gain accordingly.

The road map for further development of the MM technique leads into uncharted territory. The improvement of energy resolution has highest priority, for both the dispersive and ToF instruments. The present results indicate that values in the few-meV range are possible (even without the ToF-hybrid mode), if the unfocused He lamp is replaced by a focused one or a high-resolution Synchrotron beamline with small photon spot. Next, the reduction of the analyzed micro-area on the sample surface is a high-priority issue. In the MM family of instruments this region is defined by the field aperture, independent of the photon beam size, and can be viewed in PEEM mode. For the present electron optics regions well below 1 µm should be possible (see Fig. 2 in [1]; [122]), which can be controlled using the PEEM mode. Further, a field-free sample region would be highly desirable for many samples (non-planar surfaces, micro- and nanostructures, cleaved microcrystals). Electron-optical solutions for lenses with large angular acceptance and zero extractor field exist already [2,5]. Further, a higher multi-hit capability of the detector would strongly increase the total recording efficiency, especially for ultrafast pump-probe experiments using MMs [36-46]. Efficient multi-hit separation would be advantageous for the ToF-hybrid mode at high count rates, in order to avoid intermixing of events in the tilted data arrays. Highly parallelized DLD architectures (128 and 256 parallel channels) are under development [96]. Last not least, the instrument will be complemented by an imaging spin filter, exploiting ToF parallel recording for efficiency enhancement in the spin-filtered imaging branch [65].

**Data Availability Statement**
The data that support the findings of this study are available from the corresponding author upon reasonable request.

**Acknowledgements**
We are very grateful to Tien-Lin Lee (Diamond Light Source, Didcot, UK) and Ralph Claessen (University of Würzburg, Germany) for numerous fruitful discussions since this joint project was launched in 2016 and for a critical reading of the manuscript. Sincere thanks are also due to Oliver Schaff, Thorsten Kampen, Sven Maehl (SPECS GmbH, Berlin), Christian Tusche (Forschungszentrum Jülich), Aimo Winkelmann (presently University of Science and Technology, Krakow, Poland) and Andreas Oelsner and his team (Surface Concept GmbH, Mainz) for valuable discussions and comments on the manuscript. We gratefully acknowledge SPECS GmbH and Surface Concept GmbH for making the first prototypes of the large hemispherical analyzer for k-microscopy and the new delay-line detector with high time resolution available for the present experiment and Sergey Chernov (presently Stony Brook University, USA) for help with data evaluation. The development was funded by Deutsche Forschungsgemeinschaft (projects Scho341/16-1, Cl124/16-1 and Transregio SFB 173 Spin+X 268565370, project A02) and BMBF (project 05K19UM2).